# A Novel Hybrid Approach Using SMS and ROCOF for Islanding Detection of Inverter-Based DGs

Shahrokh Akhlaghi, *Student Member, IEEE,* Morteza Sarailoo, *Student Member, IEEE,*
Arash Akhlaghi, *Student Member, IEEE,* Ali Asghar Ghadimi, *Member, IEEE*

*Abstract*— This paper proposes a novel hybrid approach for islanding detection of inverter-based distributed generations (DG) based on combination of the Slip mode frequency-shift (SMS) as an active and rate of change of frequency (ROCOF) relay and over/under frequency relay as passive methods. This approach is utilized to force the DG to lose its stable operation and drift the frequency out of the allowed range of the frequency threshold. Performance of the proposed approach is evaluated under the IEEE 1547, UL 1741 and multiple-DG operation. The simulation results demonstrate the effectiveness of the proposed approach for detection of islanding, especially for loads with high quality factor. It operates accurately under the condition of load switching and does not interfere with the power system operation during normal condition. In other words, not only it holds the benefits of both SMS and ROCOF, but also it removes their drawbacks by having less non-detection zone and faster response.

## I. NOMENCLATURE

| | |
|---|---|
| $DG$ | Distributed Generation |
| $SMS$ | Slip Mode frequency Shift |
| $ROCOF$ | Rate of change of frequency |
| $IDM$ | Islanding detection method |
| $DC$ | Direct current |
| $NDZ$ | Non-detection zone |
| $PCC$ | Point of common coupling |
| $OFP$ | Over frequency protection |
| $UFP$ | Under frequency protection |
| $CB$ | Circuit breaker |
| $R_g, L_g$ | Resistance and inductance of power system line |
| $R, L, C$ | Load's resistance, inductance and capacitance |
| $L_{filter}$ | Inductance of the inverter filter |
| $k_P, k_I$ | Proportional and integral control factors |
| $\theta_{SMS}$ | Phase angle for SMS method |
| $\theta_m$ | Maximum phase angle in degree |
| $f_g$ | Grid frequency |
| $f_m$ | Frequency at which $\theta_m$ occurs |
| $f_{is}$ | Islanding frequency |
| $Q_f$ | Quality factor |
| $I_d, I_q$ | $d-q$ components of the current |
| $V_d, V_q$ | $d-q$ components of the voltage |
| $I_{dref}, I_{qref}$ | Reference current of $d$-$q$ components |
| $V_{dref}, V_{qref}$ | Reference voltage of $d$-$q$ components |
| $SPWM$ | Sinusoidal Pulse width modulation |
| $m$ | Modulation index amplitude |
| $\varphi$ | phase angle |

## II. INTRODUCTION

Nowadays because of increasing electrical power demand the utilization of distributed generations (DGs), e.g. wind power and photovoltaic system, is inevitable. The connection of DGs to a power system will arise many challenges. Islanding detection is one of the most important challenges over the year, which can be divided into intentional and unintentional islanding. Micro grid must be able to supply the loads during both grid connected and islanding modes [1]-[3]. According to the IEEE standards 1547, unintentional islanding refers to a condition in which part of a power system composed of DGs and loads are isolated from the rest of the system and still energized [1]. Unintentional islanding is an undesirable operation mode for the power system, and may impose some considerable problems i.e. threat to the DGs and costumer's devices, power quality problem such as voltage and frequency deviation and safety hazards to utility's staff and costumers. Thus, islanding situation must be detected immediately. Readers may refer to [4] and [5] for more details about the effects of islanding on the system performance and the reasons of islanding occurrence.

In the past decade, various methods for islanding detection have been proposed by researchers. Generally, islanding detection methods (IDM) can be classified into three main categories: communication based methods, active methods and passive islanding detection methods. Communication based methods which are based on transferring data between the DG unit and the electric utility, are reliable and effective. But, these approaches are very costly and not efficient [6].

The passive IDMs are based the observing system parameters and monitoring their changes e.g. voltage and frequency deviation, phase shift and etc.… at the point of common coupling (PCC) [6]. Passive IDMs are simple to implement and are very cost effective. Also, they are known as islanding approaches which cause no disturbance in the system. The main drawback of these approaches is their large non-detection zone (NDZ). If there is a small power mismatch between the DG and the load, the parameters deviations will not be significant which can go beyond the thresholds. Therefore, the island-mode will not be detected. Some of the passive methods are over/under frequency protection, and over/under voltage protection [7]. Voltage phase jump detection, rate of change of active power [8], rate of change of frequency protection [9], and voltage and power factor changes [10], voltage impedance and total harmonic distortion (THD) [11].

The active IDMs which are on the categories of the local approaches, injects a small disturbance on the DG's terminal at PCC to create changes in the system parameters i.e. voltage, frequency and phase. In the grid-connected mode, the small disturbance cannot make a significant changes. However, in the island-mode, the small disturbance will have significant change in system parameters. Therefore, the active methods have a smaller NDZ than passive methods. The main drawbacks of active methods are that, they have undesirable impact on power quality, due to disturbance injection. Also, Active methods have a lower speed of detection in compare with passive methods [12]. Some commonly used active techniques are Active frequency drift (AFD), Sandia frequency shift (SFS) [13], Sandia voltage shift (SVS) [14], Slip mode frequency shift (SMS) [15] and voltage phase angle difference [16].

Both active and passive methods have their own advantageous and drawbacks. Combining these two categories in order to get benefits from their advantageous and overcome the short coming of both passive and active methods, results in a new category of IDM named as hybrid approaches. Some of the recent hybrid approaches are: A hybrid islanding detection technique using voltage unbalance and frequency set point which combine the positive feedback technique as an active approach with voltage unbalance and THD as a passive was introduced in [17]-[18]. Akhlaghi et al. [15] proposed a hybrid approach based on the combination of the SMS and Q-f droop curve (active) with over/under frequency protection (passive). In [19] and [13] a novel hybrid approach combination of the SFS and Q-f droop curve was proposed.

This paper proposes a novel and robust hybrid islanding detection approach (**HIDA**) consists of an active method and passive methods. The utilized active method in this paper is the SMS method and the passive methods are the ROCOF relay and over/under frequency relay. The SMS is utilized to destabilize the inverter-based DG in the absence of the electrical network. The SMS uses positive feedback for the destabilizing purpose. The ROCOF relay is responsible to activate the SMS only when an islanded condition is suspected. To evaluate the effectiveness of the proposed **HIDA**, a simple system combination of load, DG and electric network equipped with proposed approach, and its performance is evaluated under the UL-1741 [22], IEEE 1547 [1]. Also the performance of the proposed **HIDA** is studied under the different load and multiple-DG operation condition.

The rest of paper is organized as follows. Section III describes the IEEE case study system and constant-current controller model. The SFS, ROCOF and proposed **HIDA** are presented in Section IV. Section V presents the case studies. Conclusions are drawn in Section VI.

## III. IEEE CASE STUDY SYSTEM

In this paper the proposed **HIDA** is applied to the standard IEEE case study system for islanding detection study. The case study system is shown in Fig. 1. This system includes a 100 kW inverter-based DG, connected to an electrical networked through a low pass filer, and a *RLC* local load. The electrical network is modeled as a constant voltage source behind an impedance, denoted by $L_s$ and $L_g$ in the Fig. 1. The local *RLC* load is connected to the point of common coupling (PCC). The CB is a circuit breaker [20]. As it can be seen the low pass filter is modeled as an inductance $L_{filter}$ [1]. Parameters of the system are given in TABLE I. The load parameters can be calculated as follows:

$$R = \frac{V_{pcc}^2}{P} \ , \ L = \frac{V_{pcc}^2}{2\pi f Q_f P} \ , \ Q = \frac{Q_f P}{2\pi f V_{pcc}^2} \qquad (1)$$

Where the $V_{PCC}$ is the three phase voltages at the PCC, $f$ denote the frequency, $Q_f$ is the load quality factor, $P$ and $Q$ are the load active and reactive power, respectively.

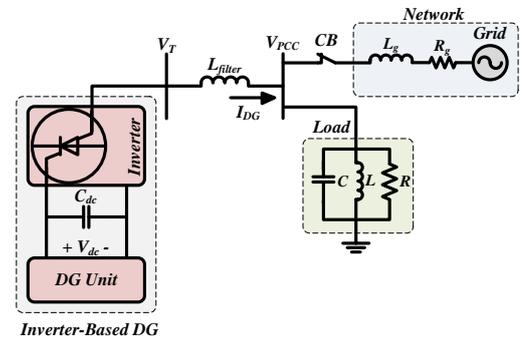

Fig. 1. IEEE case study system

TABLE I  INVERTER, SYSTEM, LOAD AND DG PARAMETER

| Inverter Parameters | |
|---|---|
| Switching frequency | 8000 Hz |
| Input DC Voltage | 900 V |
| Filter Inductance | 2.1 mH |
| Voltage (Line to Line) | 480 V |
| DG Output Active Power | 100 kW |
| **Grid Parameters** | |
| Frequency | 60 Hz |
| Grid Inductance | 0.3056 mH |
| Grid Resistance | 0.012 Ω |
| **Load Parameters** | |
| Resistance | 2.304 Ω |
| Inductance | 0.00345 mH |
| Capacitance | 2037 μF |

### A. Constant-Current Controller Model

It is assumed that the DG is designed to operate as a constant-current source with the capability of supporting its local load without any supports from the grid [1]. The model of a constant-current controller of a DG is shown in Fig. 2, which requires two types of measurements, namely, three phase output currents of the DG ($I_{DG}$) and three phase voltages at the PCC ($V_{PCC}$). The $I_{DG}$ and $V_{PCC}$ are transformed into the d-q components using the park transformation. In Fig. 2, the d-q components of currents are denoted by $I_d$ and $I_q$ and d-q components of voltages are denoted by $V_d$ and $V_q$. The $I_{dref}$ controls the active power supplied by the DG, while $I_{qref}$ controls the reactive power output of the DG. With $I_{qref}$ set equal to zero, no reactive power is supplied by the DG and the DG operates at unity power factor. Then by passing these reference currents through the phase angle transformation $I^*_{dref}$ and $I^*_{qref}$ are obtained as in (2) [22]-[24].

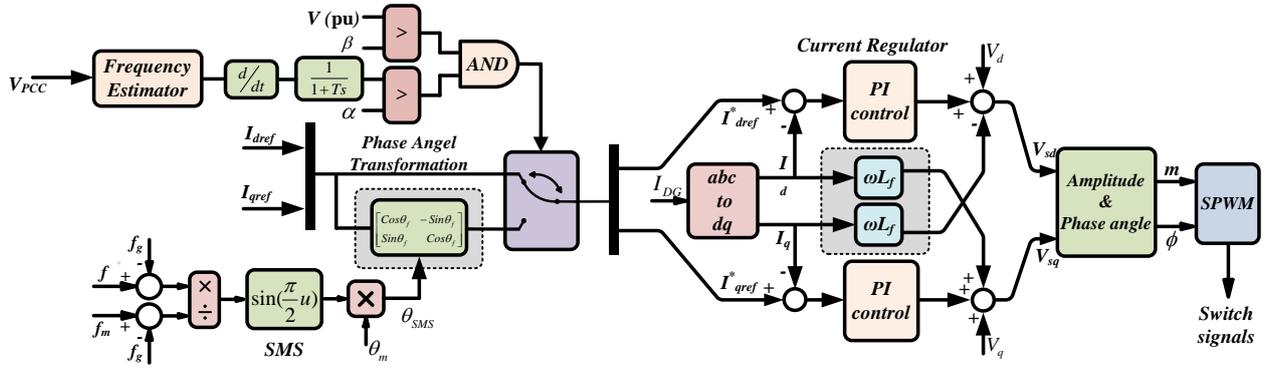

Fig. 2.  Block diagram of proposed hybrid islanding detection approach

$$\begin{bmatrix} I^*_{dref} \\ I^*_{qref} \end{bmatrix} = \begin{bmatrix} Cos\theta_f & -Sin\theta_f \\ Sin\theta_f & Cos\theta_f \end{bmatrix} \begin{bmatrix} I_{dref} \\ I_{qref} \end{bmatrix} \quad (2)$$

Next, $I^*_{dref}$ and $I^*_{qref}$ are compared with $I_d$ and $I_q$, which are the inverter output currents and pass through current regulator to the outputs $V_{dref}$ and $V_{qref}$. Given the equation (3), $V_{dref}$ and $V_{qref}$ transform to $V_{sd}$ and $V_{sq}$.

$$\begin{cases} V_{sd} = V_{dref} + V_d - L_f \omega I_q \\ V_{sq} = V_{qref} + V_q + L_f \omega I_d \end{cases} \quad (3)$$

where $L_f$ is the filter inductance. The inverter terminal voltages $V_{sd}$ and $V_{sq}$ are used to calculate the modulation index amplitude (m) and phase angle (φ) which are calculated from equations (4) and (5). SPWM is implemented to determine the on and off signals of the inverter switches. It is also possible in order to have a more robust controller a model predictive-based approach to be used [24]-[27]. This type of interface can control the DG active and reactive power output.

$$m = \sqrt{V^2_{dref} + V^2_{qref}} \quad (4)$$

$$\phi = \tan^{-1} \frac{V_{qref}}{V_{dref}} \quad (5)$$

IV. PROPOSED HYBRID ISLANDING DETECTION APPROACH

Both active and passive methods have some drawbacks. For example, passive methods have large non-detection zone [3]. In return, active methods have undesirable impact on power quality of the system [4]. The combination of these two methods, not only possible to get the benefit from their advantages; it make it possible to overcome the short coming of both passive and active methods. The proposed **HIDA** consists of SMS method as an active and ROCOF method as a passive method. SMS method, ROCOF method and proposed **HIDA** are described in details as follows:

*A. Slip Mode frequency-Shift (SMS)*

In SMS IDM a positive feedback is applied from the phase of the voltage at PCC to the phase angle of the output currents of the inverter. This causes a phase shift in the phase of the system and consequently a short-term frequency deviation [6]. The output currents of the inverter are controlled as a sinusoidal function of the frequency deviation of the voltage at PCC with respect to the network frequency before islanding. Thus, the inverter output currents can be expressed as in (6) [6].

$$i_k = \sqrt{2} I \sin[2 p f_{vk-1} t + q_{SMS}] \quad (6)$$

in which $f_{vk-1}$ is the frequency of voltage at PCC in the previous cycle and $\theta_{SMS}$ is the phase angle shift for SMS method. $\theta_{SMS}$ is given in (7) [2]:

$$\theta_{SMS} = \theta_m \sin\left(\frac{\pi}{2} \frac{f - f_g}{f_m - f_g}\right) \quad (7)$$

where $\theta_m$ and $f_m$ are the maximum phase angle in degree and the frequency at which the maximum phase angle occurs, respectively. $f_g$ is the grid frequency which is 60 Hz in this study. The application of the SMS IDM is shown in Fig. 2.

The NDZ of the SMS IDM in the $Q_f$ - $f_0$ space using the phase criteria can be expressed as follows [6]:

$$\phi_{Load} = \tan^{-1}\left(R \frac{1-\omega^2 LC}{\omega L}\right) = \tan^{-1}\left[Q_f \left(\frac{f_0}{f_{is}} - \frac{f_{is}}{f_0}\right)\right] \quad (8)$$

Where the $f_0$ is the load's resonance frequency.

$$f_0^2 + \frac{f_{is} \tan[\theta_{SMS}(f_{is})]}{Q_f} f_0 - f_{is}^2 = 0 \quad (9)$$

Equation (9) states that the NDZ of the SMS IDM depends on $\theta_m$, $Q_f$ and $f_0$. According to [6], in order to eliminate the NDZ, the $\theta_m = 25º$, destabilize the DG's frequency after the islanding for load's with high quality factor.

### B. Rate of Change of Frequency

ROCOF relay is a passive IDM. In the case of occurrence of an islanding it is expected that the frequency of voltage at PCC faces a high frequency rate change (*df/dt*). The ROCOF relay is used to detect islanding mode based on the computed rate of change of frequency over a few cycles (2~50 cycles). The measured signals are pass through a low-pass filter, to eliminate the high-frequency transient made by power system components. The ROCOF relay is activated only when the rate of frequency change and the voltage phasor at PCC both are greater than a pre-specified threshold for rate of frequency change and a minimum voltage. The voltage condition is imposed to avoid a ROCOF relay malfunction during a long electrical motor startup or short circuit fault. The application of the ROFOC relay is shown in Fig. 2. The low-pass filter output is compared with pre-defined setting ($\alpha$). In the case that the *df/dt* is greater than $\alpha$, a trip signal will be send to the circuit breaker. To avoid ROCOF relay's malfunction, the terminal voltage is compared with the minimum adjustable voltage setting ($\beta$). When the terminal voltage is reduced to less than $\beta$, the ROCOF relay's trip signal will be inactive [28]-[29]. The ROCOF relay setting considering the 60 Hz nominal frequency is in the range of 0.1 ~ 1.2 Hz/s for $\alpha$. In this study, regarding the fact that the maximum frequency deviation of the system under study is about 1 Hz/s, thereupon, $\alpha$ is set to be 1.2 Hz/s. $\beta$ is also set to be 0.85 p.u, to helping the proposed **HIDA** not confuse the motor start up condition with islanding occurrence, due to the fact that in motor start up the condition the terminal voltage may reduce.

### C. The proposed HIDA

The proposed **HIDA**, is the combination of the SMS IDM as an active method and ROCOF relay as a passive method. The SMS is utilized to destabilize the inverter-based DG in the absence of the electrical network. The SMS uses positive feedback for the destabilizing purpose. The ROCOF relay is responsible to activate the SMS only when an islanded condition is suspected. By margining this two IDM, the islanding detection performance can be improved. The block diagram of the proposed approach is shown in Fig. 2. The frequency of the system is estimated and deviation of the frequency is calculated and passes through the low-pass filter. If the *df/dt* is greater than the threshold, a trip signal will be sent to a multiple switch to activate the SMS IDM. Otherwise, pre-defined $I_{dref}$ and $I_{qref}$ will be injected. The major problem of ROCOF relay is that, when the load and generation inside the island are closely matched, detection of islanding is difficult. Therefore, the adjustment of the ROCOF parameters should be taken into account. If the thresholds set to be too low, it may result to unnecessary trip of DG; and if the thresholds are set too high, islanding may not be detected. In the proposed approach, if the islanding suspected, the ROCOF relay will send a trip signal to activate the SMS IDM. Then SMS IDM will detects the islanding.

It should be noted that, the kay point about merging SMS and ROCOF is keeping the ROCOF relay thresholds as low as possible, to make the islanding detection possible even with closely match capacity of load and DG. According to [6], the performance of the SMS IDM depends on $\theta_m$. Larger value of $\theta_m$ may expedite the islanding detection procedure. On the other hand, larger value of $\theta_m$ reduces the power quality. Regarding the fact that the SMS IDM is not active continuously, the system power quality will significantly improve. As a result, in this study, $\theta_m$ is set to be equal to 25º.

## V. SIMULATION RESULTS

In order to assess the proposed **HIDA**, it is applied to the IEEE case study system, shown in Fig. 1 with the given parameters in the TABLE I under different operating conditions. It is shown that the proposed approach complies with IEEE 1547 standard and UL 1741 procedures. Also, the performance of the proposed approach is studied under different load conditions, load switching and various load quality factor.

### A. UL-1741 Testing Condition

The performance of the proposed method **HIDA** under UL-1741 test conditions are evaluated based on the given values in TABLE II and TABLE III. It is assumed that the DG is equipped with the proposed **HIDA** and islanding is occurred at t = 2 s. Based on the UL 1741 an acceptable anti-islanding method should be able to withstands local load active powers of 50%, 100% and 125% of the rated power of the DG, given in TABLE II

TABLE II      LOAD PARAMETERS FOR ACTIVE POWER MISMATCH

| %P | %Q | R(Ω) | L (H) | C (μF) | $Q_f$ | $f_r$ |
|---|---|---|---|---|---|---|
| 100 | 100 | 2.304 | 0.00345 | 2037 | 1.77 | 60.05 |
| 50 | 100 | 4.603 | 0.00345 | 2037 | 3.54 | 60.05 |
| 125 | 100 | 1.841 | 0.00345 | 2037 | 1.41 | 60.05 |

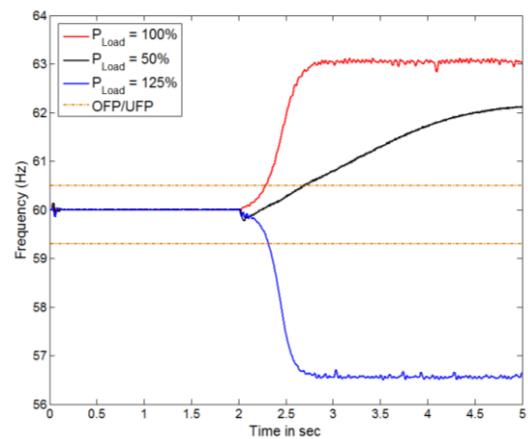

Fig. 3. PCC frequency with the proposed **HIDA** for active power mismatch

The corresponding results of scenario 1 (TABLE II), are shown in Fig. 3. From this figure it is clear that, during normal operation condition, the DG is stable and the corresponding frequency is fixed at the nominal value. While the islanding occurs at *t* = 2 s, DG loses its stable operation and voltage frequency deviates from over/under frequency thresholds. To evaluate the effectiveness of the proposed **HIDA**, simulation results of system without proposed **HIDA** is represented in Fig. 4. From this figure, it can be seen that in a system without proposed **HIDA**, the voltage frequency after islanding occurrence could not exceeds the over/under frequency thresholds. According to UL 1741 Std, changes in the reactive power of load should be considered between 95% to 105% of the nominal power [22]. The considered cases in this paper are tabulated in TABLE III. The frequency of the voltage at the PCC for system equipped with the proposed approach for cases correspond to TABLE III are shown in Fig. 5. Fig. 5 shows that for loads with different reactive power the proposed **HIDA** is able to detect the islanding within a short time period after the islanding occurrence time.

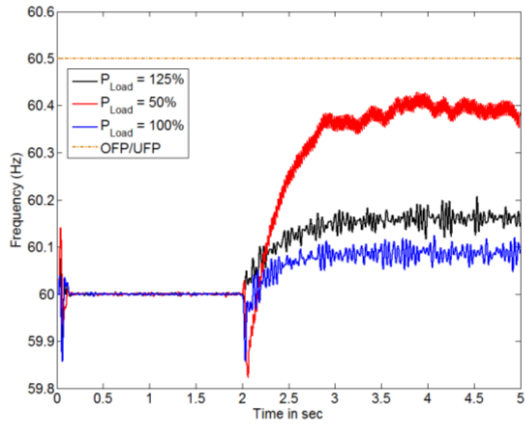

Fig. 4. PCC frequency with the SMS IDM method for active power mismatch

TABLE III     LOAD PARAMETERS FOR UL 1741 TESTING

| Cases | %P | %Q | R (Ω) | L (H) | C(µF) | $Q_f$ | $f_r$ |
|---|---|---|---|---|---|---|---|
| 1 | 100 | 105 | 2.304 | 0.003278 | 2037 | 1.81 | 61.6 |
| 2 | 100 | 102 | 2.304 | 0.003381 | 2037 | 1.78 | 60.6 |
| 3 | 100 | 101 | 2.304 | 0.003419 | 2037 | 1.77 | 60.3 |
| 4 | 100 | 100 | 2.304 | 0.00345 | 2037 | 1.77 | 60 |
| 5 | 100 | 99 | 2.304 | 0.003488 | 2037 | 1.76 | 59.7 |
| 6 | 100 | 98 | 2.304 | 0.003519 | 2037 | 1.76 | 59.7 |
| 7 | 100 | 95 | 2.304 | 0.003623 | 2037 | 1.73 | 58.6 |

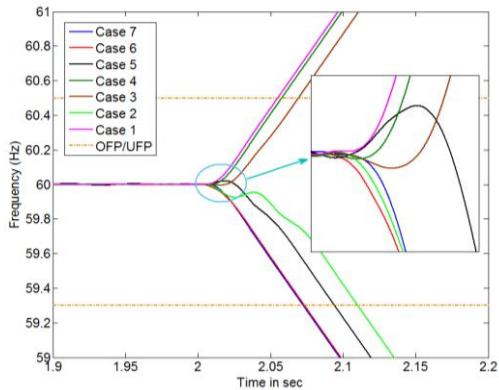

Fig. 5. PCC frequency with the proposed method for different reactive power values of the load.

### B. Effect of Load Switching

Here the proposed **HIDA** is assessed under load switching condition in the grid-connected operation mode. The purpose of the test is to ensure that the proposed **HIDA** is not sensitive to load switching. In other word, the proposed approach should be able to distinguish between the islanding condition and load switching condition. It is assumed that the new load is connected to the PCC at *t* = 2 s and disconnected at *t* = 3 s. Three different considered scenarios are as follows:

- Load active power of 100 kVA and power factor of 0.8 lead.
- Load active power of 100 kVA and power factor of 1.0.
- Load active power of 100 kVA and power factor of 0.8 lag.

The frequency and the voltage at the PCC for the above mentioned scenarios with system equipped with the proposed approach are shown in Fig. 6. From this figure, it is clear that, the proposed approach does not have considerable effect on the voltage and frequency of the system during normal operation condition. In other word, the proposed approach is not sensitive to load switching.

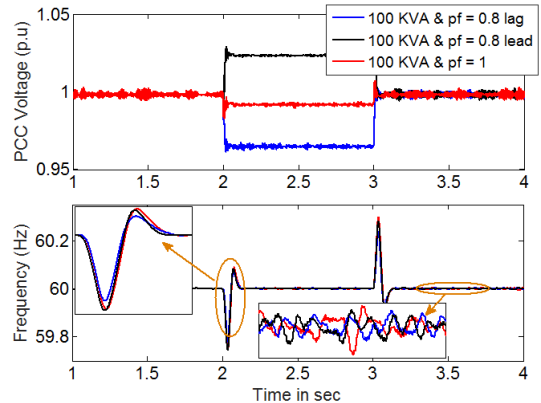

Fig. 6. PCC voltage and frequency behavior during a load switching event.

### C. Influence of Load Quality Factor

Here the performance of the proposed **HIDA** is evaluated for loads with different quality factors. There are different standard in order to test the islanding detection methods for considering the impact of load quality factor. For example, IEEE Standard strictly requires that any acceptable islanding detection method should detect the islanding in less than 2s after the event occurrence time for RLC loads with the $Q_f \leq 2.5$ [1]. However, UL 1741 Std, requires the load with $Q_f \leq 1.8$ [22]. In this work, in order to show the efficiency of the proposed approach, the load quality factor is adjusted to different values in the range of 0.5 to 8.1 by changing the load inductance and capacitance. For all the cases studied in this section it is assumed that the load inductance and capacitance are adjusted in a way to keep the inverter load active power at 100% of its output, and the load reactive power at 100% of the balanced condition, respectively. These cases are summarized in TABLE IV and corresponding simulation results for a system with and without proposed **HIDA** are depicted in Fig. 7 and Fig. 8, respectively.

TABLE IV    LOAD PARAMETERS FOR DIFFERENT VALUES OF $Q_F$

| R (Ω) | L (H) | C (μF) | Qf | fr |
|---|---|---|---|---|
| 2.304 | 0.0122 | 575.4 | 0.5 | 60.07 |
| 2.304 | 0.0061 | 1150 | 1 | 60.1 |
| 2.304 | 0.00345 | 2037 | 1.77 | 60 |
| 2.304 | 0.00203 | 3452 | 3 | 60.12 |
| 2.304 | 0.00145 | 4850 | 4.21 | 60 |
| 2.304 | 0.000957 | 7350 | 6.38 | 60 |
| 2.304 | 0.000754 | 9330 | 8.1 | 60 |

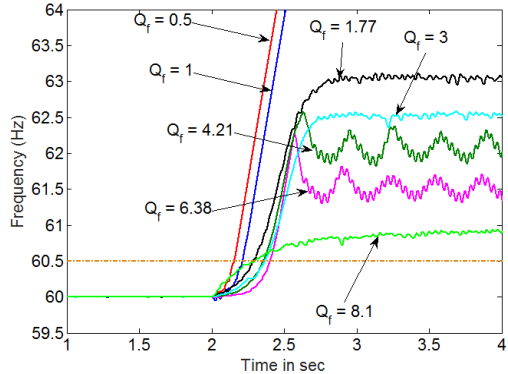

Fig. 7.    Frequency at PCC for different quality factors with proposed **HIDA**

From Fig. 7, it can be seen that, the proposed **HIDA** deviates the frequency from over/under frequency thresholds in small period of time for a load with high quality factors. In other words, the proposed **HIDA** is able to detect the islanding in a short period of time for a broad range of load quality factors considered by different standards. To evaluate the efficiency of the proposed approach, it is compared with the SMS IDM with $\theta_m = 15°$. As it was mentioned earlier, increasing the value of $\theta_m$ may degrade the power quality of the system. While increasing $\theta_m$ in the proposed approach, because of ROCOF relay, has a relatively low impact on the system power quality. From Fig. 8, it is clear that, for load with high quality factor (i.e. $Q_f \geq 1.8$) the frequency of the voltage at the PCC does not deviate from the frequency relay thresholds. In other word, the islanding condition may not be detected for load with high quality factor using the SMS IDM solely. As a result, it can be concluded that the proposed **HIDA** has a faster speed of response and smaller NDZ in comparison with SMS IDM.

VI.    ONCLUSION

This paper proposed a new hybrid islanding detection approach as a combination of the SMS IDM as an active method and ROCOF relay and over/under frequency relay as passive methods. The proposed approach works in a way that, the ROCOF relay measures the rate of change of frequency, if the islanding condition suspected, the signal will be sent to activate the SMS IDM to detect the islanding situation. The kay point in proposed approach, is to keep the ROCPF relay threshold as low as possible to catch the frequency deviation even for cases that have closely match between load and DG and on the other hand to adjust the SMS IDM's parameters to decrease the NDZ and detection time.

The proposed approach has been studied under different conditions such as load switching and loads with high quality factor. Also, it has been evaluated under various tests such as IEEE 1547 and UL 1741. From the simulation studies it can be concluded that, the proposed approach is accurate and fast enough to detect the islanding occurrence. Also, the proposed approach has a smaller NDZ than the SMS IDM. Last but not least, the proposed approach has relatively low negative effect on the system power quality.

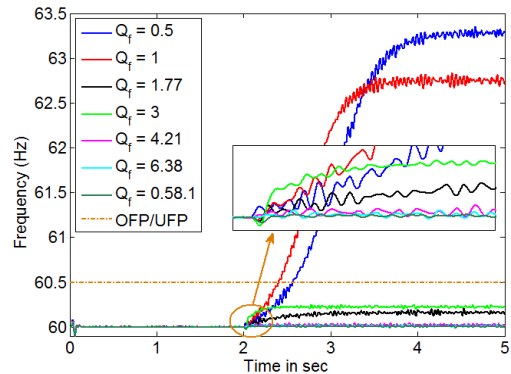

Fig. 8.    Frequency at PCC for different quality factors without proposed **HIDA**

## VIII. BIOGRAPHIES

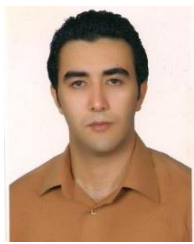

**Shahrokh Akhlaghi** (S'14) received the M.S. degree in electrical engineering from the Amirkabir University of Technology, Tafresh, Iran, in 2010. He is currently working toward the Ph.D. degree in electrical engineering from Binghamton University, State University of New York (SUNY), Binghamton, NY, USA. He was a Lecturer in Department of Electrical and Computer Engineering at Tehran North Branch of Islamic Azad University, Tehran, Iran from 2010 to 2013.

His research interests include power system dynamics, state estimation, phasor measurement unit, Kalman filtering, distributed generation and islanding detection of DGs. Shahrokh Akhalghi is the chairman of the IEEE PES Student Branch Chapter and IEEE Young Professional affinity group at Binghamton.

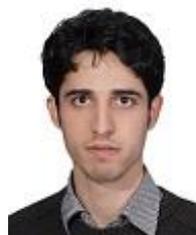

**Morteza Sarailoo** (S'14) received the B.Sc. and M.S. degrees in electrical engineering from the Mazandaran University, Babolsar, Iran, in 2009 and Babol University of Technology, Babol, Iran, in 2012, respectively. He currently is a Ph.D candidate at the Electrical and Computer Engineering Department of the Binghamton University, Binghamton, NY, USA.

His research interests include power system transient stability, state estimation, renewable energy, and control systems. Morteza sarailoo is a member of the IEEE Power and Energy Society (PES).

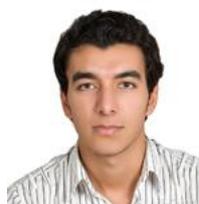

**Arash Akhlaghi** received the B.Sc. degree from Islamic Azad University, Saveh, Iran, in 2010. He is currently working toward the M.S. degree in electrical engineering from Leibniz University Hannover, Hannover, Germany.

His research interests include Distributed Generation Islanding Detection, Renewable Energy, Power Systems and Power Quality. Arash Akhlaghi is a member of the IEEE Power and Energy Society (PES).

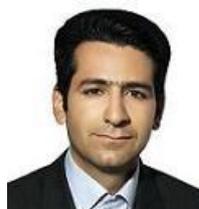

**Ali Asghar Ghadimi** received his M.Sc and Ph.D. from AmirKabir University of Technology (Tehran Polytechnic), Tehran, IRAN in 2002 and 2008 respectively in Power Engineering. Currently, he is Assistant Professor in Department of Electrical Engineering at Arak University, Arak, IRAN since 2008.

His research interests are in the area of Micro Grid, Distributed Generation, Fuel Cell and Simulation and Analysis of Power System.